\documentclass{ws-ijmpb}
\usepackage{epsfig}

\def\avg#1{\langle#1\rangle}

\def\be{\begin{equation}}       \def\ee{\end{equation}}
\def\bea{\begin{eqnarray}}      \def\eea{\end{eqnarray}}
\def\PRA{Phys. Rev. A~}
\def\PRB{Phys. Rev. B~}
\def\PRD{Phys. Rev. D~}
\def\RMP{Rev. Mod. Phys. ~}
\def\PRL{Phys. Rev. Lett.~}
\def\nn{\nonumber}

\begin{document}

%\markboth{Congjun Wu}{Unconventional  Bose-Einstein Condensations
%Beyond the ``No-node'' Theorem}

\catchline{}{}{}{}{}
%
%%%%%%%%%%%%%%%%%%%%%%%%%%%%%%%%%%%%%%%%%%%%%%%%%%%%%%%%%%%%%%%%%%%%

\title{Quintet pairing and non-Abelian vortex string in
spin-$\frac{3}{2}$ cold atomic systems
%\footnote{For the title, try not to 
%use more than 3 lines. Typeset the title in 10 pt 
%Times roman, uppercase and boldface.}
}

\author{\footnotesize Congjun Wu
%\footnote{Typeset names in 
%10~pt Times roman, uppercase. Use the footnote to indicate 
%the present or permanent address of the author.}
}

\address{ Department of Physics, University 
of California, San Diego \\  La Jolla, CA 92093-0319, USA. \\
wucj@physics.ucsd.edu}

\author{Jiangping Hu}
\address{
Department of Physics, Purdue University, West Lafayette, IN 47907, USA
}
\author{Shou-Cheng Zhang}
\address{Department of Physics, McCullough Building, Stanford
University, Stanford CA 94305-4045, USA}

\maketitle

\begin{history}
\received{(Day Month Year)}
\revised{(Day Month Year)}
\end{history}

\begin{abstract}
We study the $s$-wave quintet Cooper pairing phase ($S_{pair}=2$)
in spin-3/2 cold atomic systems and identify various novel
features which do not appear in spin-1/2 pairing systems.
A single quantum vortex is shown to be energetically less stable
than a pair of half-quantum vortices. 
The half-quantum vortex exhibits the global analogue of the non-Abelian
Alice string and $SO(4)$ Cheshire charge in gauge theories. The
non-Abelian half-quantum vortex loop enables topological generation of
quantum entanglement.
\end{abstract}


\section{Introduction}
Optical traps and lattices open up a whole new direction in the
study of strongly correlated large spin systems by using cold atoms
with hyperfine multiplets. In spin-1 bosonic systems (e.g.
$^{23}$Na and $^{87}$Rb), spinor condensations, spin textures and
nematic orders have generated a great deal of attention
\cite{myatt1997,ho1998,mueller2003,reijnders2004,demler2002,zhou2001,zhou2003}. On the other hand, large spin fermions
also exhibit many exciting novel features. For instance, the
multi-particle clustering instability, {\it i.e.}, a multi-particle
counterpart of Cooper paring, is not allowed in spin 1/2 systems
due to Pauli's exclusion principle, but is possible in large spin
systems \cite{wu2005,stepanenko1999,kamei2005}. Furthermore, large spin
fermions offer a unique playground to study high symmetries which
do not appear in usual condensed matter systems.  
We have proved that spin-3/2 fermionic systems with contact interactions,
which can be realized by atoms such as $^{132}$Cs, $^9$Be,
$^{135}$Ba, $^{137}$Ba and $^{201}$Hg, enjoy a generic $SO(5)$
symmetry  for the continuum model with $s$-wave scattering interactions and the 
lattice Hubbard model, whose exactness is regardless of dimensionality, 
lattice geometry and external potentials \cite{wu2003,wu2006}.
Such a high symmetry without fine-tuning is rare in both condensed 
matter and cold atom many-body systems.
The important consequences of this symmetry are systematically
investigated, including the protected degeneracy in collective excitations, 
the absence of the quantum Monte-Carlo sign problem,
the four-fermion quartetting superfluidity, and even stronger 
quantum magnetic fluctuations than spin-1/2 systems
\cite{chern2004,chen2005,controzzi2005,hattori2005,wu2005,xu2008,wu2009},
which has been summarized in a review article \cite{wu2006}.

On the other hand, important progress has been made in the
fault-tolerant topological quantum computation \cite{freedman2002,kitaev2003,kitaev2006,freedman2004}.
The key idea is that by using non-Abelian statistics in two
dimensions, particles can be entangled in a robust way against
local disturbances. The promising candidate systems to implement
topological quantum computation include the non-Abelian quantum
Hall states with fermions at the filling $\nu=5/2$
\cite{freedman2005} and bosons at $\nu=1$ \cite{fradkin1998}, and
also the $p_x+ip_y$ pairing state of spinless fermions 
\cite{read2000,stone2005}. 
In this paper, we
show that due to the $SO(5)$ symmetry in spin-3/2 cold atomic
systems \cite{wu2003,wu2005}, the $s$-wave quintet Cooper
pairing state ($S_{pair}=2$) in such systems provides another
opportunity to topologically generate quantum entanglement between
the particle and the non-Abelian half-quantum vortex (half-quantum vortex) loop.

The half-quantum vortex in superfluids with the spin degree of freedom is an exotic 
topological defect as a global analogue of the Alice string in 
gauge theories \cite{schwarz1982,march1992,bucher1999,striet2003}. 
The half-quantum vortex loop can possess spin quantum number 
which is an example of the Cheshire charge phenomenon.
An Abelian version of the global Alice string and Cheshire charge
exists in the triplet superfluid of the $^3$He-A phase
\cite{volovik2000,khazan1985,saloma1985,saloma1987,mcgraw1994}, 
where the spin $SU(2)$ symmetry is
broken into the $U(1)$ symmetry around the $z$-axis. A
remarkable property is that both quasi-particles and spin wave
excitations reverse the sign of their spin quantum numbers $s_z$ when
going through the half-quantum vortex loop. Meanwhile the half-quantum vortex loop also changes $s_z$
to maintain spin conservation. However, due to the
Abelian nature of this $U(1)$ Cheshire charge, no entanglement is
generated in this process.

In this article, we investigate the non-Abelian Alice string and
the topological generation of quantum entanglement through the
non-Abelian Cheshire charge in spin-3/2 systems. The quintet
Cooper pairing order parameters in the polar basis
form a 5-vector of the $SO(5)$ symmetry group. 
The ground state exhibits the polar condensation where the
$SO(5)$ symmetry is broken into $SO(4)$ \cite{wu2003}.
This allows the half-quantum vortex loop to possess the non-Abelian $SO(4)$ Cheshire
charge, in contrast to the $U(1)$ Cheshire charge in the
$^3$He-A phase. We also explore the high symmetry effects on
collective spin excitations and the structure of the half-quantum vortex line as a
$\pi$-disclination in the spin channel. We show that
by driving the fermion quasiparticle (or spin-wave impulse) through
the half-quantum vortex loop, quantum entanglement between them
is topologically generated.
This effect has a potential application in the  topological quantum
computation.

\section{Quintet Pairing}
The spin-$\frac{3}{2}$ system is the simplest one to support the 
$s$-wave quintet pairing with total spin $S_T=2$.
In such a system with contact interactions, a hidden $SO(5)$ symmetry 
\cite{wu2003} 
arises as follows: the four-component spinor 
\bea \psi=(c_{\frac{3}{2}},
c_{\frac{1}{2}},c_{-\frac{1}{2}}, c_{-\frac{3}{2}})^T
\eea
forms the spinor representations of the $SU(4)$ group which is the unitary
transformation of the four-component spinor. 
Each of the four component contributes the same to the kinetic energy
which is explicitly $SU(4)$ symmetric.
Generally speaking, the interactions break this symmetry to a lower
level.
From the view of the spin $SU(2)$ group, the interactions can be classified
into the total spin 0(singlet), 1 (triplet), 2 (quintet), 3(septet) channels.
For the contact interactions, say, s-wave scattering,  only total spin
singlet and quintet channel are allowed as required by the Pauli's  
exclusion principle.
Interestingly, the spin $SU(2)$ singlet channel can also be interpreted 
as an $SO(5)$ singlet, and the spin $SU(2)$ quintet channel can be interpreted
as an $SO(5)$ vector channel.
This $SO(5)$ group only lies in the particle-hole channel as a subgroup
of the $SU(4)$ group.
Thus, the remaining symmetry with the $s$-wave scattering interaction
is $SO(5)$ without fine-tuning of parameters.

We denote the interaction strength in the spin singlet and triplet
channels as $g_0$ and $g_2$, respectively.
We consider the case of $g_2<0$ where the quintet channel Cooper pairing 
dominates, and further neglect the interaction in the singlet channel. The
mean field  Hamiltonian reads
\bea
H_{MF}&=& \int d^D r \Big\{ \sum_{\alpha=\pm
\frac{3}{2}, \pm \frac{1}{2}}
 \psi^\dagger_\alpha({ r})
\big (\frac{-\hbar^2\nabla^2}{2M}-\mu\big) \psi_\alpha({ r})
+\sum_{a=1\sim 5} \chi^\dagger_a(r) \Delta_a(r) +h.c.\nn \\
&-&\frac{1}{g_2} \Delta_a^*(r)\Delta_a(r)\Big\},
\label{eq:mfhm} \eea with $D$ the spatial dimension, $\mu$ the
chemical potential, and $M$ the atom mass. $\Delta_a$ is 
proportional to the
ground state expectation value of the quintet pairing operators
$\chi_a$ by 
\bea
\Delta_a(r)= g_2\avg{\chi_a(r)},  
\eea
where $a=1\sim 5$.
The five $\chi_a$ operators are the spin channel
counterparts of the five atomic $d$-orbitals ($d_{xy}, d_{xz},
d_{yz}, d_{3z^2-r^2}, d_{x^2-y^2}$), and
transform as a 5-vector under the $SO(5)$ group. Explicitly,
they are expressed as \bea 
\chi^\dagger_a(r)= -\frac{i}{2}
\psi^\dagger_\alpha(r) (\Gamma^a R)_{\alpha\beta}
\psi^\dagger_\beta(r), \eea 
where the five $4\times4$ Dirac
$\Gamma^a ~(a=1\sim5)$ matrices are defined as
\bea 
\Gamma^1&=&\left (
\begin{array} {cc}
0 & -i I\\
i I& 0
\end{array} \right), \ \ \
\Gamma^{2,3,4}=\left ( \begin{array}{cc}
{\vec \sigma}& 0\\
0& {-\vec \sigma} \end{array}\right), \ \ \
\Gamma^5=\left( \begin{array} {cc}
0& I \\
I & 0 \end{array} \right ), \nonumber
\eea
which satisfy the anti-commutation relation as 
$\{\Gamma^a, \Gamma^b\}= 2\delta_{ab}$,
and $R=\Gamma^1\Gamma^3$ is the charge conjugation matrix
\cite{wu2003}.

\begin{figure}
\centering\epsfig{file=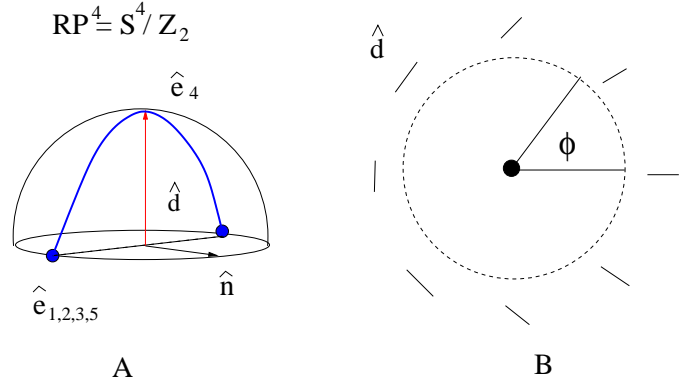,clip=1,width=0.7\linewidth,angle=0}
\caption{A) The Goldstone manifold of $\hat d$ is a 5D
hemisphere $RP^4$. It contains a class of non-contractible loops
as marked by the solid curve. B) The $\pi$-disclination of $\hat
d$ as a half-quantum vortex. Assume that $\hat d \parallel \hat e_4$  at
$\phi=0$.
As the azimuthal angle $\phi$ goes from 0 to $2\pi$,
$\hat d$ is  rotated at
the angle of $\phi/2$ around any axis $\hat n$ in the $S^3$
equator spanned by $\hat e_{1,2,3,5}$. } \label{fig:GSManifold}
\end{figure}

This $SO(5)$ symmetry leads to new interesting results about the
pairing structure in the ground state
and the corresponding Goldstone  modes. Within the BCS
theory, Ref. \cite{ho1999} showed that the ground state of Eq.
\ref{eq:mfhm} is an $SO(3)$ polar condensate without noticing the
hidden $SO(5)$ symmetry. We conclude here that the ground state is
generically an $SO(5)$ polar condensate. The order parameters can
be parameterized as $\Delta_a= |\Delta| e^{i\theta} d_a$, where
$\theta$ is the $U(1)$ phase, $\hat d=d_a \hat e_a$ is a 5$D$
unit vector, and $\hat e_a (a=1\sim5)$ form a set of basis for the
internal spin space. Rigorously speaking, $\hat d$ is a
directionless director instead of a true vector because
$\Delta_a$'s contain a $Z_2$ gauge symmetry of \bea \hat d
\rightarrow -\hat d, \ \ \, \theta\rightarrow \theta+ \pi. \eea
Thus the Golstone manifold is $[SO(5)
\otimes U(1)]/[SO(4)\otimes Z_2] =RP^4 \otimes U(1)$, where $RP^4$
is a 5$D$ hemisphere instead of an entire $S^4$ sphere as
depicted in Fig. \ref{fig:GSManifold} A.

The general Ginzburg-Landau free energy without the gradient term
for quintet pairing has been given in Ref. \cite{ho1999} with
three independent quartic terms.
In the special spin-$\frac{3}{2}$ case with the SO(5) symmetry,
it can be simplified into a more convenient SO(5) invariant 
form with only two quartic terms
\bea
F_{GL}&=&\int d^3 r  \Big\{
\gamma \vec \nabla \Delta^*_a . \vec \nabla \Delta_a  
+ \alpha(T)  \Delta^*_a  \Delta_a
+\frac{\beta_1}{2}  (\Delta^*_a  \Delta_a)^2
+ \frac{\beta_2}{2} \sum_{1\le a<b\le 5}  L_{ab}^2 \Big \}, \ \ \
\eea
where 
\bea
L_{ab}=  \frac{\Delta^*_a \Delta_b -\Delta^*_b \Delta_a}{\sqrt 2}.
\eea
The first quartic term describe the density interaction among Cooper pairs
while the second one describes the spin interaction among them.
The polar-like state is favorable to make $L_{ab}=0$ at   $\beta_2>0$,
while the axial state is favorable at $\beta_2<0$.
Around the critical temperature $T_c$, the parameters $\alpha, \beta_{1,2}$
was  calculated from the Gor'kov expansion 
in Ref. \cite{ho1999}
\bea
\alpha&=& -\frac{1}{2} \frac{dn}{d\epsilon} (1- \frac{T}{T_c}),\ \ \
\beta_1=\beta_2= \frac{1}{2} \frac{dn}{d\epsilon} 
\frac{7 \zeta(3)}{8 \pi^2  T_c^2},
\eea
where $n$ is the particle density, 
$dn/d \epsilon$ is the density of states at the Fermi level, 
and $\zeta(x)$ is the Riemann zeta function.
The coefficient of $\gamma$ can also be calculated as
\bea
\gamma= \frac{n \hbar^2}{4M} \frac{7 \zeta(3)}{8 \pi^2  T_c^2}.
\eea

\section{Gross-Pitaevskii equation}
At zero temperature, the Ginzburg-Landau free energy fails.
The low energy degree of freedom is described by an effective
Gross-Pitaevskii equation for the Cooper pairs.
In addition to the usual phonon mode, four
branches of spin wave modes carrying the spin quantum number
$S=2$ arise because of the
spontaneous symmetry breaking from $SO(5)$ to $SO(4)$.
In other words, they can be called ``spin nematic waves''. 
For small fluctuations, the spin wave modes decouple from the phase mode.
For the purpose of describing collective excitations, Cooper pairs
can be treated as composite bosons. This treatment gives a good
approximation to the phonon mode in the neutral singlet BCS
superfluid \cite{aitchison1995,stone1995}.

Here we generalize this method to the quintet pairing by using a
phenomenological Hamiltonian for spin-2 bosons \bea H_{eff}&=&\int
d^D r \Big\{ \frac{\hbar^2}{4 M} \sum_{1\le a\le 5} \nabla
\Psi^\dagger_a  \nabla \Psi_a
+\frac{1}{2\chi_\rho} (\Psi^\dagger_a \Psi_a 
-\rho_0)^2  \nonumber \\
&+&\frac{1}{2\chi_{sp}} \sum_{1\le a<b\le 5}(\Psi^\dagger_c L^{ab}_{cd} \Psi_d)^2
\Big\},
\label{eq:effham}
\eea
where $\Psi_{a}$'s are the boson operators in the polar basis,
the equilibrium Cooper pair density $\rho_0$ is half of
the particle density $\rho_f$,
$\chi_\rho$ and $\chi_{sp}$ are proportional to the compressibility
and $SO(5)$ spin susceptibility respectively.
We define  the $SO(5)$ generators in the $5\times 5$ vector
representation as
\bea
L^{ab}_{cd}=i(\delta_{ac}\delta_{bd}-\delta_{ad}\delta_{bc}).
\label{eq:vecso5}
\eea

The Landau parameters in the $SO(5)$ symmetric Fermi liquid theory
for spin-$\frac{3}{2}$ systems can be decomposed into three sectors 
as $SO(5)$ scalar $F^s_l$ (density), vector $F^v_l$(spin-quadrapole
density), tensor $F^t_l$ (spin and spin-octupole densities),
where $l$ denotes the quantum number of orbital angular momentum
\cite{wu2003,wu2006}.
Taking into account the Fermi liquid correction,
\bea
\chi_\rho=\frac{N_f}{4(1+F_0^s)}, \ \ \, \chi_{sp}=\frac{N_f}{4(1+F_0^t)},
\eea
where $N_f$ is the fermion density of states at the Fermi energy.

We introduce $\rho(r)$ as the Cooper pair density and
$l_{ab}(r)$ as the  $SO(5)$ spin density, and parameterize
$\Psi_a=\sqrt\rho_0 e^{i\theta} d_a$.
Using the standard commutation rules between
$l_{ab}$ and  $\hat d_a$, $\rho$ and $\theta$,
we arrive at
\bea
&&\partial_t l_{ab}= \frac{\hbar^2\rho_{sp}}{2 M} ( d_a \nabla^2  d_b
-  d_b \nabla^2  d_a),~ 
\chi_{sp}\partial_t  d_a = - l_{ab}  d_b, \nn \\
&&\chi_\rho \partial_t^2 \theta -\frac{\hbar^2\rho_s}{2M} \nabla^2 \theta
=0, \ \ \
\label {EOM1}
\eea
where $\rho_{sp}$ is the superfluid  density
and $\rho_{s}$ is the spin superfluid density.
At $T=0K$ in a Galilean invariant system, $\rho_s$ is just
$\rho_f/2$, while $\rho_{sp}$ receives Fermi liquid
corrections \cite{leggett1975} as 
\bea
\frac{\rho_{sp}}{\rho_s}=\frac{1+F^v_1/3}{1+ F^s_1/3}
\eea
where $F_1^v$ is the Landau parameter
in the $SO(5)$ vector channel \cite{wu2003}.
The spin wave and sound velocities are obtained
as
\bea
v_{sp}=\sqrt{\frac{\rho_0}{2 \chi_{sp} M}}, \ \ \,
v_s=\sqrt{\frac{\rho_0}{2 \chi_\rho M}},
\eea
respectively.

\section{Half-quantum vortex}
The fundamental group of the GS manifold is
\bea
\pi_1(RP^4\otimes U(1))=Z\otimes Z_2.
\eea
The $Z_2$ feature 
gives rise to the existence of the half-quantum vortex as a stable topological 
defect as depicted in
Fig. \ref{fig:GSManifold} B. As we move along a loop enclosing the
half-quantum vortex, the $\pi$ phase mismatch in the $\theta$ field is offset by a
$\pi$-disclination in the $d$-field, thus $\Delta_a$'s are maintained
single-valued. Energetically, a single quantum vortex is less
favorable than a pair of half-quantum vortices. From Eq. \ref{eq:effham}, the
static energy function can be written as 
\bea 
E=\int d^D r\frac{\hbar^2}{4M} \Big\{ \rho_s (\nabla \theta)^2+ \rho_{sp}
(\nabla \hat d)^2 \Big \}. 
\eea 
The energy density per unit length
of a single quantum vortex is $E_1=\frac{\hbar^2}{4M} \rho_s
\log \frac{L}{a}$, while that of two isolated half-quantum vortices
is $E_2=\frac{\hbar^2}{8M} (\rho_s+\rho_{sp}) \log
\frac{L}{a}$. Although at the bare level $\rho_s=\rho_{sp}$,
$\rho_{sp}$ receives considerable Fermi liquid correction and
strong reduction due to quantum fluctuations in the 5D internal
space. Generally speaking, the relation,
\bea
\rho_{sp}<\rho_{s},
\eea
holds in terms of their renormalized values. 
Then a single quantum vortex is
fractionalized into a pair of half-quantum vortices. In the presence of rotation,
the half-quantum vortex lattice should appear instead of the usual single quantum
vortex lattice. As a result of the doubling of vortex numbers,
their vortex lattice constants differ by a factor of $\sqrt 2$.

The half-quantum vortex was also predicted in the  $^{3}$He-A phase, where 
$\hat d$  is a 3D vector  defined for spin-1 Cooper pairs. However, the
dipole locking effect favors the $d$-vector aligned along the
fixed direction of the $l$-vector, {\it i.e.}, the direction of the
$p$-wave orbital angular momentum. As a result, the two half-quantum vortices are
linearly confined by a string of the mismatched $d$ and
$l$-vectors. In contrast, the orbital part of the quintet pairing
is $s$-wave, no dipole locking effect exists.

\begin{figure}
\centering\epsfig{file=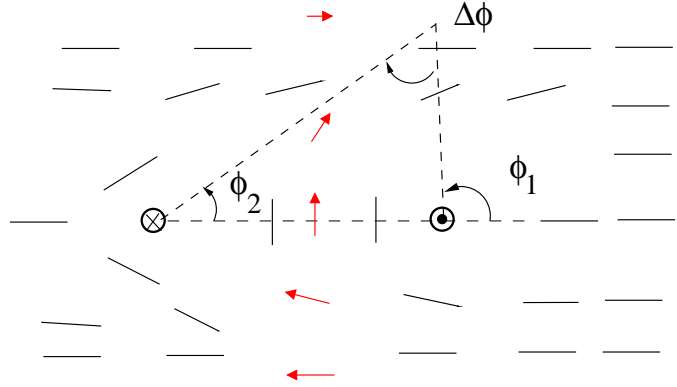,clip=1,width=0.7\linewidth,angle=0}
\caption{The configuration of a $\pi$-disclination pair or loop
described by Eq. \ref{eq:hqvpair}.
$\phi_{1,2}$ and $\Delta \phi$ are azimuthal angles and
$\hat d (\vec r) \parallel \hat e_4$
as $\vec r\rightarrow \infty$.
After a fermion passes the half-quantum vortex loop, the components with $S_z=\pm\frac{3}{2}$
change to $S_z=\pm\frac{1}{2}$ and {\it vice versa} with
an $SU(2)$ matrix defined in Eq. \ref{eq:spflip}.
}\label{fig:halfpair}
\end{figure}

\section{SO(4) Cheshire charge}
The single half-quantum vortex line behaves like the Alice string because
a quasi-particle changes its spin quantum number after it
adiabatically moves around the half-quantum vortex once.
For example, in the $^3$He-A phase, a quasi-particle with spin $\uparrow$
flips its spin to $\downarrow$ up to a $U(1)$ Berry phase.
The half-quantum vortex in the quintet superfluid  behaves as a non-Abelian
generalization with the $SU(2)$ Berry phase.
Without loss of generality, we assume that
$\hat d$ is parallel to $\hat e_4 $ at the azimuthal angle $\phi=0$.
As $\phi$ changes from $0$ to $2\pi$, $\hat d$ is rotated
at the angle of $\phi/2$ in the plane
spanned by $\hat e_4$ and  $\hat n$, where $\hat n$
is a unit vector perpendicular to $\hat e_4$, {\it i.e.}, a vector
located in the $S^3$ sphere spanned by $\hat e_{1,2,3,5}$. 
We define such a rotation operator as $U(\hat n, \phi/2)$. When $U$
acts on an $SO(5)$ spinor, it takes the form of 
\bea
U(\hat n,\frac{\phi}{2})=
\exp\{ -i \frac{\phi}{2} \frac{ n_b \Gamma^{b4}}{2} \}
\eea where
$\Gamma^{b4}=i[\Gamma^b,  \Gamma^4]/2$ are $SO(5)$ generators in
the $4\times 4$ spinor representation; 
when $U$ acts on an $SO(5)$ vector, it behaves as 
\bea
U(\hat n,\phi/2)=\exp\{ -i \frac{\phi}{2} n_b L^{b4} \}
\eea where $L^{ab}$'s are the $SO(5)$
generators in the $5\times 5$ vector representation
explicitly defined in Eq. \ref{eq:vecso5}. The resulting
configuration of $\hat d$ is
\bea
\hat d (\hat n, \phi)=  U(\hat n, \phi/2)
\hat d (\hat n, 0)
= \cos \frac{\phi}{2} \hat e_{4} -\sin
\frac{\phi}{2} \hat n .
\eea
As fermionic quasi-particles circumscribe around the vortex line adiabatically,
at $\phi=2\pi$ fermions with $S_z=\pm\frac{3}{2}$ are rotated into
$S_z=\pm\frac{1}{2}$ and {\it vice versa}.
For convenience, we change the basis $\Psi$ for the fermion wavefunction
to $(|\frac{3}{2}\rangle,|-\frac{3}{2}\rangle,|\frac{1}{2}\rangle,
|-\frac{1}{2}\rangle)^T$.
After taking into account the $\pi$ phase winding of $\theta$,
$\Psi$ transforms by
\bea
\Psi_a\rightarrow\Psi^{\prime}_a = i U(\hat n,
\pi)_{\alpha\beta} \Psi_{\beta}= \left(
\begin{array}{cc}
0&W\\
W^\dagger&0
\end{array}
\right)_{\alpha\beta} \Psi_\beta
\label{eq:spflip}
\eea
where $W$ is an $SU(2)$ Berry phase depending on the direction of $\hat n$
on the $S^3$ sphere as
\bea
W(\hat n)=\left (\begin{array}{cc}
n_3+in_2&-n_1-in_5\\
n_1-in_5& n_3-in_2
\end{array}
\right). \eea
The non-conservation of spin in this adiabatic
process is not surprising because the $SO(5)$ symmetry is
completely broken in the configuration depicted in
Fig. \ref{fig:GSManifold} B.

A more interesting but related concept is the Cheshire charge,
which means that a pair of the half-quantum vortex loop can carry $SO(4)$ spin quantum numbers.
An intersection between the half-quantum vortex loop and a perpendicular plane is
depicted in Fig. \ref{fig:halfpair}, where $\phi_{1,2}$ are
respect to the vortex and anti-vortex cores
respectively. Without loss of generality, we assume $\hat d(\vec r)
\rightarrow \hat e_4$ as $r \rightarrow \infty$ where an
$SO(4)$ symmetry generated by $\Gamma_{ab}(a,b=1,2,3,5)$ is
preserved. In analogy to Fig. \ref{fig:GSManifold} B, the $\hat d$
vector is described by the difference between two azimuthal angles
$\Delta\phi=\phi_2-\phi_1$ as \bea \hat d (\hat n, \Delta \phi) =
\cos \frac{\Delta \phi}{2} \hat e_{4} -\sin \frac{\Delta \phi}{2}
\hat n,  \label{eq:hqvpair} \eea where $\hat n$ again is a unit
vector on the $S^3$ equator. This classical configuration is
called a phase-sharp state denoted as $|\hat n\rangle_{vt}$.
Because the above $SO(4)$ symmetry is only broken within a small
region around the half-quantum vortex loop, quantum fluctuations of $\hat n$
dynamically restore the $SO(4)$ symmetry as described by the
Hamiltonian \bea H_{rot}&=& \sum_{a,b=1,2,3,5} \frac{M^2_{ab}}{2
I},~ M_{ab}= i (\hat n_a
\partial_{\hat n_b}-\hat n_b \partial_{ \hat n_a}), \ \ \ \eea
with the moment of inertial
\bea
I= \chi_{sp} \int d^D r ~ \rho_0
\sin^2 \frac{\Delta \phi}{2}.
\eea
Thus the zero modes $|\hat
n\rangle_{vt}$ are quantized into the global $SO(4)$ Cheshire
charge states, which are a non-Abelian generalization of the $U(1)$
case in the $^3$He-A phase  \cite{mcgraw1994}. 
The global Cheshire charge density is localized around the half-quantum vortex loop. In
contrast, the Cheshire charge in gauge theories is non-localized
\cite{striet2003}.

The $SO(4)$ algebra can be grouped into two commutable sets of
$SU(2)$ generators as 
\bea
T_1
(T^\prime_1)&=&\frac{1}{4}(\pm\Gamma_{35}-\Gamma_{12}), 
\ \ \
T_2
(T^\prime_2) = \frac{1}{4}(\pm\Gamma_{31}-\Gamma_{25}),  \nonumber \\
T_3(T^\prime_3)&=&\frac{1}{4}(\pm\Gamma_{23}-\Gamma_{15}).
\eea
$T_{1,2,3}$
and $T^\prime_{1,2,3}$ act in the subspaces spanned by $|\pm
\frac{3}{2}\rangle$ and $|\pm \frac{1}{2}\rangle$, respectively.
$SO(4)$ representations are denoted by $|T, T_3; T^{\prime },
T_3^\prime \rangle$, {\it i.e.},  the direct-product of representations
of two $SU(2)$ groups. The half-quantum vortex loop in the $SO(4)$ Cheshire charge
eigenstates is defined as 
\bea
|T T_3; T^\prime
T_3^\prime\rangle_{vt} =\int_{\hat n \in S^3} d \hat n~ F_{T T_3;
T^\prime T_3^\prime} (\hat n) ~|\hat n\rangle_{vt}, 
\eea
 where $ F_{T
T_3; T^\prime T_3^\prime} (\hat n)$ are the $S^3$ sphere harmonic
functions. 
%The $|\hat n\rangle_{vt}$ state is the non-Abelian
%generalization of the familiar phase-sharp state in $U(1)$
%theories, while 
Thus $|T T_3; T^\prime T_3^\prime\rangle_{vt}$ is the
non-Abelian generalization of the usual number-sharp state in
$U(1)$ theories.

\begin{figure}
\centering\epsfig{file=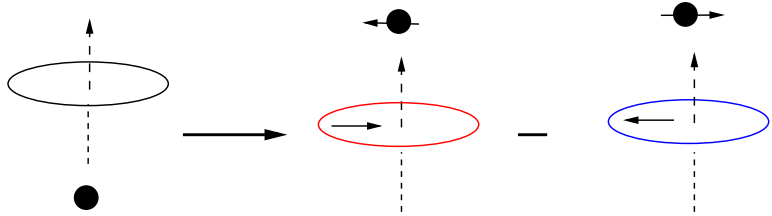,clip=1,width=0.8\linewidth,angle=0}
\caption{The topological generation of quantum entanglement.
The initial state $|i\rangle$ is a product state of a zero charged 
half-quantum vortex loop and a quasiparticle with $S_z=3/2$, both of which are a singlet of
the $SU(2)$ group of $T^\prime$  acting in the subspace spanned by 
$|\pm \frac{1}{2}\rangle$.
The final state $|f\rangle$ is an EPR state of the $T^\prime$ group
as described by Eq. \ref{eq:entag}.
} \label{fig:entangle}
\end{figure}

\section{Generation of entanglement}
When a particle passes the half-quantum vortex loop, $\Delta \phi$ changes from
$0$ to $2\pi$. The conservation of the $SO(4)$ spin is ensured by
exciting the Cheshire charges and generating quantum entanglement
between the particle and the half-quantum vortex loop. 
We demonstrate this process explicitly through a concrete example, 
with the initial state $|i\rangle$ made up from a zero charged half-quantum vortex loop 
and a quasiparticle with  $S_z=\frac{3}{2}$ as
\bea |i\rangle &=&\int_{\hat n\in S^3} d\hat n  ~|\hat
n\rangle_{vt} \otimes (u~ c^\dagger_{\frac{3}{2}} +v
~c_{-\frac{3}{2}})|\Omega\rangle_{qp}, \eea where $|\Omega\rangle_{qp}$ is
the vacuum for Bogoliubov particles. 
For each phase-sharp state
$|\hat n\rangle_{vt}$, the particle changes spin according to Eq.
\ref{eq:spflip} in the final state $|f\rangle$. The superposition of
the non-Abelian phase gives 
\bea |f\rangle &=& \int_{\hat n\in S^3}d
\hat n ~ \Big\{ u ~(W^\dagger_{11} c^\dagger_{\frac{1}{2}}+
W^\dagger_{21}
c^\dagger_{-\frac{1}{2}})  + v~ (W^T_{12} c_{\frac{1}{2}}
+  W^T_{22} c_{-\frac{1}{2}} )\Big\} \ \ \
|\hat n \rangle_{vt} \otimes |\Omega\rangle_{qp}
\nonumber \\
&=&\int_{\hat n\in S^3}d \hat n  (\hat n_3-i \hat
n_2) |\hat n \rangle_{vt} \otimes (u~ c^\dagger_{\frac{1}{2}} +v
~c_{-\frac{1}{2}})|\Omega\rangle_{qp}  \nonumber \\
&-&\int_{\hat n\in S^3}d \hat n (\hat n_1-i \hat
n_5) |\hat n\rangle_{vt} \otimes (u~ c^\dagger_{-\frac{1}{2}} -v
~c_{\frac{1}{2}})|\Omega\rangle_{qp}, \ \ \
\label{eq:cheshire} 
\eea 
as depicted in Fig. \ref{fig:entangle}. In terms of the $SO(4)$
quantum numbers, $|i\rangle$ is  a product state of
$|00;00\rangle_{vt} \otimes
|\frac{1}{2}\frac{1}{2};00\rangle_{qp}$, and $|f\rangle$ is
{\small
\bea
|\frac{1}{2}\frac{1}{2}; \frac{1}{2}\frac{-1}{2}
\rangle_{vt}\otimes |00;\frac{1}{2}\frac{1}{2} \rangle_{qp}-
|\frac{1}{2}\frac{1}{2}; \frac{1}{2}\frac{1}{2}\rangle_{vt}
\otimes |00;\frac{1}{2}\frac{-1}{2}\rangle_{qp}. \ \ \
\label{eq:entag}
\eea}In the channel of $(T^{\prime },T_3^\prime)$,
the final state is exactly an entangled Einstein-Podolsky-Rosen 
(EPR) pair made up from the half-quantum vortex loop and the quasi-particle. 
We note that this mechanism of generating the
quantum entanglement is entirely topological, dependent only on
whether the trajectory of the quasi-particle lies inside or
outside of the half-quantum vortex loop. In contrast, 
the half-quantum vortex loop in $^3$He-A
system only exhibits the $U(1)$ Cheshire charge, thus the final
state is still a product state without the generation of 
entanglement.

Similarly, the entanglement between a spin wave impulse and 
the half-quantum vortex loop can also be generated.
We consider the four local transverse bases $\hat e_b
(b=1,2,3,5)$ at $\Delta \phi=0$. Assume that the initial state 
made up from
the half-quantum vortex loop and the spin wave is 
\bea
|i^\prime \rangle=
\int_{\hat n\in S^3} d \hat n~ |\hat n\rangle_{vt} \otimes (\hat
e_1+ i\hat e_5)_{sw},
\eea  where spin wave impulse
carries $S_z=2$. For each phase-sharp state $|\hat n \rangle_{vt}$ of
the half-quantum vortex, the frame bases at $\Delta \phi=2\pi$ transform to $\hat
e_a\rightarrow \hat e_a - 2 \hat n (\hat e_a \cdot \hat n)$. Thus
the entanglement is generated in the final state $|f^\prime\rangle$
as 
\bea
|f^\prime\rangle&=&\int_{\hat n\in S^3} d \hat n~ |\hat
n\rangle_{vt} \otimes (\hat e_1 + i\hat e_5)_{sw}
\nonumber \\
&-&2 \int_{\hat n\in S^3} d \hat n~  ( n_1 + i n_5) n_b
|\hat n\rangle_{vt}\otimes \hat e_{b,sw}.
\eea

\section{Conclusion}
Recently, Bose condensation of the $^{174}$Yb atom and sympathetic
cooling between $^{174}$Yb and the fermionic atom of $^{171}$Yb
\cite{takasu2003} have been achieved. Their electron
configurations are the same as the Ba atoms except an  inside full-filled
$4f$ shell, thus the spin-3/2 systems of $^{135}$Ba and $^{137}$Ba
can be possibly realized in the near future. At the present time,
scattering lengths of these two Ba atoms are not available.
However, considering the rapid developments in this field, we are
optimistic about the realization of the quintet pairing state and
the associated non-Abelian topological defects.

We briefly discuss here the factors that limit the life time of 
the entanglement which come from spin decoherence.
As shown above, the generation of entanglement only depends 
on whether the particle trajectory penetrates the
half-quantum vortex loop or not, but does not on the detail of how
it penetrates the vortex loop.
Thus this process is topological and is robust. 
However, spin decoherence does come from the interaction between
particles and vortex loops with the bulk
low energy collective excitations.
The quintet pairing states have gapless spin-wave excitations.
The particle spin and the Chesire charge of the half-quantum 
vortex loop can flip when spin-waves scatter with them, which
is the leading order spin decoherence mechanism.
Nevertheless, spin-waves are Goldstone particles which only
interact with other excitations through the derivative coupling,
{\it i.e.}, the coupling constant vanishes at long wave length limit.
At low temperatures, only long wavelength spin waves are excited,
thus their spin decoherence effect is small.

In summary, we have studied the quintet pairing state in 
spin 3/2 fermionic systems with the $SO(5)$ symmetry, 
including its Goldstone modes and the non-Abelian topological
defects. The non-Abelian Berry phase effect and
the Cheshire charge behavior are analyzed in detail. The topological
mechanism of generating the quantum entanglement between
quasi-particles and the half-quantum vortex loop could be useful for topological
quantum computation.

\section*{Acknowledgments}

C.W. thanks E. Fradkin and J. Slingerland for helpful discussions.
This work is supported by the US NSF Grant No. DMR-9814289,
and the US Department of Energy, Office of Basic Energy Sciences
under contract DE-AC03-76SF00515. C.W. is supported by the the
US NSF Grant No. DMR-0804775.

%\bibliographystyle{prsty}
%\bibliography{orbital,exciton,spin32}

\end{document}